\documentclass{article}
\usepackage{spconf,amsmath,graphicx}
\usepackage{blindtext}
\usepackage{lipsum}
\usepackage{float}
\usepackage{enumitem}
\usepackage{wrapfig}
\usepackage{graphicx}
\usepackage{xpatch}
\usepackage[dvipsnames]{xcolor}
\usepackage{booktabs}
\usepackage{makecell}
\usepackage{multirow}
\usepackage{tikz}
\DeclareRobustCommand\solid {\tikz[baseline=-0.6ex]\draw[thick] (0,0)--(0.5,0);}
\DeclareRobustCommand\dotted{\tikz[baseline=-0.6ex]\draw[thick,dotted] (0,0)--(0.54,0);}
\DeclareRobustCommand\dashed{\tikz[baseline=-0.6ex]\draw[thick,dashed] (0,0)--(0.54,0);}
\DeclareRobustCommand\dashdot {\tikz[baseline=-0.6ex]\draw[thick,dash dot] (0,0)--(0.5,0);}

\newcommand\blfootnote[1]{%
  \begingroup
  \renewcommand\thefootnote{}\footnote{#1}%
  \addtocounter{footnote}{-1}%
  \endgroup
}
\usepackage{hyperref}
\hypersetup{
    colorlinks=true,
    linkcolor=Mulberry,
    filecolor=purple,      
    urlcolor=CadetBlue,
    citecolor=RoyalBlue,
}

\title{STRING SOUND SYNTHESIZER ON GPU-ACCELERATED FINITE DIFFERENCE SCHEME}
\name{Jin Woo Lee $^{1,2}$\qquad Min Jun Choi $^{1,5}$\sthanks{Work done during an internship at Music and Audio Research Group (MARG), Seoul National University.}\qquad Kyogu Lee $^{1,2,3,4}$}
\address{
    $^1$MARG\quad $^2$Department of Intelligence and Information, Seoul National University\\
    $^3$AIIS \quad\
    $^4$IPAI \quad\ 
    $^5$Department of Nuclear Engineering, Seoul National University
}

\begin{document}
\ninept
\maketitle
\begin{abstract}
This paper introduces a nonlinear string sound synthesizer, based on a finite difference simulation of the dynamic behavior of strings under various excitations. The presented synthesizer features a versatile string simulation engine capable of stochastic parameterization, encompassing fundamental frequency modulation, stiffness, tension, frequency-dependent loss, and excitation control. This open-source physical model simulator not only benefits the audio signal processing community but also contributes to the burgeoning field of neural network-based audio synthesis by serving as a novel dataset construction tool. Implemented in PyTorch, this synthesizer offers flexibility, facilitating both CPU and GPU utilization, thereby enhancing its applicability as a simulator. GPU utilization expedites computation by parallelizing operations across spatial and batch dimensions, further enhancing its utility as a data generator.
\end{abstract}
\blfootnote{
AIIS and IPAI abbreviates Artificial Intelligence Institute of Seoul National University and Interdisciplinary Program in Artificial Intelligence, respectively. Project demo page: {\scriptsize\url{https://bit.ly/torch-fdtd-string}}\\
Code: {\scriptsize\url{https://github.com/jin-woo-lee/torch-fdtd-string}}
}

\vspace{-5mm}
\section{Introduction}
\label{sec: intro}
\vspace{-1mm}

Investigation of wave propagation along strings, encompassing both theoretical and experimental dimensions, has persisted for well over a century \cite{rayleigh1896theory,donkin1884acoustics}.
In the relentless pursuit of verisimilitude and expressive fidelity in simulating wave phenomena, a multitude of endeavors have arisen to bridge the gap between theoretical underpinnings and empirical sound measurements \cite{fletcher1964normal}.
More recently, developments harnessing the computational power to mimic the intricate physical processes in musical instruments have given rise to numerical sound synthesis, which has emerged as a cornerstone in the realm of virtual sound synthesis \cite{ruiz1970technique,hiller1971synthesizing}.
Diverse perspectives within the sound synthesis domain, some possessing a physical interpretation and others diverging from such constraints, have enriched the field.
However, this paper is firmly centered on physical modeling, particularly focusing on the finite difference method \cite{bilbao2009numerical,smith2010physical}.

Within the domain of physical modeling, various methodologies exist, each endowed with unique strengths that lend themselves to modeling specific target systems \cite{bilbao2004wave,valimaki2011virtual,valimaki2005discrete}.
Modal synthesis, for instance, predominantly pertains to linear problems and is adept at capturing static or non-modular configurations.
Conversely, digital waveguides, rooted in digital filter design and scattering theory, and closely aligned with wave digital filters, offer an elegant solution to mitigating computational costs, particularly for musical instruments whose vibrating components can be approximated as one-dimensional linear media described, in a first-order approximation, by the wave equation \cite{smith2010physical,fettweis1986wave}.
Meanwhile, a separate line of inquiry concentrates on scrutinizing instrument behavior, aligning model equations with empirical data, and refining instrument design.
A prevailing approach in this endeavor involves the utilization of finite difference schemes.

The finite difference scheme, a method for approximating partial differential equations (PDEs), begets mathematical models known as finite difference time domain (FDTD) methods.
Its historical origins trace back to the 1920s \cite{courant1928partiellen}, where its primary applications were found in the domains of fluid dynamics and electromagnetics simulations.
Notably, the application of FDTD to sound synthesis was posited as early as the 1970s by Ruiz \cite{ruiz1970technique}, primarily in the context of vibrating linear \cite{chaigne1994numerical,bensa2003simulation,ducasse2005waveguide} and nonlinear \cite{kirchhoff1897vorlesungen,carrier1945non} strings.
However, these models, though valuable, often neglect longitudinal motion, thereby failing to account for energy transfer between the transverse and longitudinal directions \cite{narasimha1968non}.

Compared to other physical modeling methodologies like modal synthesis, digital waveguides, or wave digital filters, FDTD distinguishes itself through its proficiency in modeling nonlinear and multi-dimensional systems.
While this approach demands substantial computational resources, recent strides in GPU-based computing have rendered it increasingly accessible \cite{bilbao2013large}.
A multitude of studies have explored the implementation of FDTD on GPUs, not only for sound synthesis \cite{bilbao2013large} but also for room acoustics \cite{southern2010finite,webb2011computing,mehra2012efficient}.

Beyond the realm of GPU-accelerated simulations, extensive research has been dedicated to more realistic sound simulation methodologies, primarily executed on CPUs \cite{bilbao2023models,bank2006physics}.
Sound synthesists have diligently endeavored to faithfully model the nonlinearities inherent in strings \cite{tolonen2000modeling,bilbao2004energy,ducceschi2022simulation}.
Simultaneously, numerical analysts have been committed to enhancing the efficiency and stability of the FDTD scheme without compromising synthesis quality \cite{ducceschi2020real,russo2022efficient,willemsen2022dynamic}.
While these efforts have yielded groundbreaking contributions to nonlinear string synthesis via finite difference methods, a notable deficiency is the lack of open-source, readily deployable implementations.
This paper, therefore, makes the following contributions:
\begin{itemize}[leftmargin=5mm]
    \item The provision of an open-source nonlinear string synthesizer.
    \item The synthesizer adeptly simulates the vibration of a planar stiff string, accommodating frequency-dependent loss and various excitations, all achieved through FDTD.
    \item The synthesizer can operate with randomly augmented control parameters and material properties, thus serving as a versatile dataset construction toolkit.
    \item The synthesizer offers flexibility by supporting both CPU and GPU computations, thereby providing parallelization options.
\end{itemize}
In the subsequent sections, the paper delves into the historical context of numerical sound synthesis, intricate details of the finite difference scheme, and a comprehensive analysis of the synthesizer.

\newpage
\begin{table}
    \centering
    \footnotesize
    \begin{tabular}{c|c|l}\toprule
        \textbf{Related to} & \textbf{Notation} & \textbf{Description}
        \\\midrule
        \multirow{5}{*}{Scheme} & $u$  & transverse displacement \\
        & $\zeta$  & longitudinal displacement \\
        & $\theta$  & implicit scheme free parameter \\
        & $k$  & temporal resolution ($k=1/f_s$) \\
        & $h$  & spatial resolution \\
        \midrule
        \multirow{8}{*}{String} & $\gamma$  & wave speed ($\gamma=L^{-1}\sqrt{T_0/\rho A}$) \\
        & $\alpha$  & stiffness-to-tension ratio ($\alpha=\sqrt{YA/T_0}$) \\
        & $Y$  & Young's modulus \\
        & $A,L$  & cross-sectional area, characteristic length \\
        & $\rho,T_0,I$  & density, nominal tension, moment of inertia \\
        & $\kappa$ & string stiffness ($\kappa=\sqrt{YI/\rho AL^4}$) \\
        & $\sigma_0,\sigma_1$ & frequency-independent, -dependent loss \\
        \midrule
        \multirow{1}{*}{Pluck} & $c_0$, $x_{\tt P}$  & plucking peak amplitude, position \\
        \midrule
        \multirow{3}{*}{Bow} & $x_{\tt B}$  & bowing position \\
        & $v_{\tt B}$  & bowing velocity \\
        & $F_{\tt B}$  & bowing force \\
        \midrule
        \multirow{7}{*}{Hammer} & $x_{\tt H}$  & hammering position \\
        & $u_{\tt H}, v_{\tt H}$  & hammer initial displacement, initial velocity \\
        & $F_{\tt H}$  & hammering force \\
        & $\mathcal{M}$  & hammer-string mass ratio \\
        & $\omega_{\tt H}$  & hammer stiffness parameter \\
        & $\alpha_{\tt H}$  & hammer nonlinear stiffness exponent \\
        \bottomrule
    \end{tabular}
    \vspace{-2mm}
    \caption{List of notations of the parameters for the simulation.}
    \label{tab: notation}
    \vspace{-3mm}
\end{table}
\section{Nonlinear String Vibration}
\label{sec: nonlinear string vibration}
\vspace{-1mm}
\setlength{\intextsep}{9pt}%
\setlength{\columnsep}{9pt}%
\begin{wrapfigure}{r}{0.23\linewidth}
    \vspace{-4mm}
    \centering
    \includegraphics[width=\linewidth]{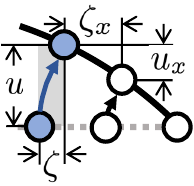}
    \vspace{-7mm}
    \caption{String coordinate system at the deflected ({\solid}) state is shown along with the equilibrium ({\color{lightgray}\dashed}) state for the reference.}
    \label{fig: planar}
    \vspace{-10mm}
\end{wrapfigure}

Kirchhoff \cite{kirchhoff1897vorlesungen} and Carrier \cite{carrier1945non} introduced a generalization of a wave equation to a nonlinear string model, where the longitudinal motion is averaged over the length of the string and only transverse motion is explicitly modeled \cite{bilbao2009numerical}.
While this approximation is valid under certain conditions \cite{bank2006physics}, perceptually important effects such as phantom partials arise from the coupling between longitudinal and transverse motion \cite{bilbao2023models}.
A model of such planar string vibration is as follows \cite{morse1986theoretical}.
\begin{subequations}\label{eqn: wave}
    \begin{align}
        u_{tt} &= \gamma^2 u_{xx} - \gamma^2(\alpha^2-1)\left(\frac{\partial\Phi}{\partial q}\right)_x\label{eqn: wave-u}\\
        \zeta_{tt} &= \gamma^2\alpha^2 \zeta_{xx} - \gamma^2(\alpha^2-1)\left(\frac{\partial\Phi}{\partial p}\right)_x\label{eqn: wave-z}
    \end{align}
\end{subequations}
Table \ref{tab: notation} summarizes the descriptions of each notation.
$u(x,t)$ and $\zeta(x,t)$ represent the transverse and longitudinal displacements of a string, respectively.
These quantities are defined for time $t\geq0$ and the spatial domain $x\in[0,L]$.
Figure \ref{fig: planar} shows an illustration of the string at spatial grid $x$ with planar displacements $u$ and $\zeta$.
By defining $q=u_x$ and $p=\zeta_x$, a general form of nonlinear string \cite{morse1986theoretical} can be rewritten as \eqref{eqn: wave} with auxiliary system $q_t=(u_t)_x$ and $p_t=(\zeta_t)_x$ \cite{bilbao2005conservative}.
The function $\Phi$, representing a contribution to the potential energy density due to the nonlinearity, is often Taylor series approximated about $u_t=\zeta_t=0$ with additional simplification \cite{bilbao2005conservative,anand1969large} as $\Phi = \frac{1}{2}q^2 - \frac{1}{2}pq^2 - \frac{1}{8}q^4$.
Under this choice of $\Phi$, equation \eqref{eqn: wave} reduces to the following set of equations.
\begin{subequations}\label{eqn: wave-truncated}
    \begin{align}
        u_{tt} &= \gamma^2 u_{xx} - \gamma^2\frac{\alpha^2-1}{2}\left(q^3+2pq\right)_x \label{eqn: wave-u-truncated}\\
        \zeta_{tt} &= \gamma^2\alpha^2 \zeta_{xx} - \gamma^2\frac{\alpha^2-1}{2}\left(q^2\right)_x \label{eqn: wave-z-truncated}
    \end{align}
\end{subequations}
To solve this nonlinear PDE, this paper adopts a finite difference time domain approach.

\section{Finite Difference Scheme}
\label{sec: simulation}
Before delving into the discretization of eq \eqref{eqn: wave-truncated}, it is imperative to establish a spatio-temporal grid.
Temporal discretization involves dividing time into discrete steps $t=nk$, where $n$ starts at 0 and $k$ is chosen as $1/f_s$ based on sample rate $f_s$.
Similarly, spatial domain $x$ is discretized into $N$ equidistant intervals of length $h$ with positions $x=lh$, where $l$ ranges from 0 to $N$.

\subsection{Numerical method}
Various schemes have been devised to approximate equation \eqref{eqn: wave-truncated} \cite{bilbao2009numerical}.
These methods demand meticulous handling to address potential issues stemming primarily from numerical errors.
As one kind of such method, the scheme employed in this study is as follows:
\begin{subequations}\label{eqn: fd}
    \begin{align}
        \delta_{tt}u &= \gamma^2\delta_{xx}^{({\tt t})} u - \gamma^2\frac{\alpha^2-1}{2} \delta_{x+}^{({\tt t})}\left(q^2\mu_{t\cdot}q+2q\mu_{tt}p\right) \label{eqn: fd-u}\\
        \delta_{tt}\zeta &= \gamma^2\alpha^2\delta_{xx}^{({\tt l})} \zeta - \gamma^2\frac{\alpha^2-1}{2}\delta_{x+}^{({\tt l})}\mathcal{I}_{({\tt t})\to ({\tt l})}\left(q\mu_{t\cdot} q\right) \label{eqn: fd-z}
    \end{align}
\end{subequations}
Here, $\delta$ and $\mu$ are the difference and averaging operators, respectively, with subscripts indicating temporal ($t$) and spatial ($x$) axes.
For instance, $t\cdot$ and $tt$ represent centered and double-temporal differences, while $x+$ and $xx$ represent forward and double-spatial differences.
Superscripts $({\tt t})$ and $({\tt l})$ denote the operator direction over transverse or longitudinal displacement grids, respectively.

The distinguishing characteristic of the scheme described in equation \eqref{eqn: fd} lies in its utilization of different spatial resolutions along the transverse and longitudinal axes.
Denote the number of spatial grid sizes in eq \eqref{eqn: fd-u} and \eqref{eqn: fd-z} by $N^{(\tt t)}$ and $N^{(\tt l)}$.
Notably, $\mathcal{I}_{({\tt l})\to({\tt t})}$ and $\mathcal{I}_{({\tt t})\to({\tt l})}$ correspond to the $p$-th order upsampling interpolant of the spatial grid and its corresponding downsampling operation, respectively.
For more exploration of the intricacies of this scheme, we encourage interested readers to refer to Bilbao \cite{bilbao2009numerical}.

\subsection{Extensions with stiffness and dissipation}
\label{ssec: stiffness and dissipation}
While tension usually takes precedence, stiffness contributes to audible inharmonicity, often sought after for its unique musical characteristics, notably apparent in instruments like pianos.
Also, actual strings manifest a notably intricate dissipation profile, marked by the non-uniform decay rates of their various frequency components.
Research on modeling strings with stiffness \cite{morse1986theoretical} and frequency-dependent dissipation \cite{bensa2003simulation} have been studied for a long time and a numerical scheme to model them can be written as follows.
\begin{subequations}\label{eqn: fd-lossy}
    \begin{align}
        (\theta + (1-\theta)\mu_{x\cdot})\delta_{tt}u &= \gamma^2\delta_{xx}^{(\tt t)} u  - \kappa^2\delta_{xxxx}^{(\tt t)}u\nonumber\\
        &+ \gamma^2\frac{\alpha^2-1}{2}\delta_{x+}^{(\tt t)}\left(q^2\mu_{t\cdot}q + 2q\mu_{tt}p\right) \nonumber\\
        &- 2\sigma_0^{(\tt t)}\delta_{t\cdot}u + 2\sigma_1^{(\tt t)}\delta_{t\cdot}\delta_{xx}^{(\tt t)}u \label{eqn: fd-u-lossy}\\
        \delta_{tt}\zeta = \gamma^2\alpha^2\delta_{xx}^{(\tt l)} \zeta 
        &+ \gamma^2\frac{\alpha^2-1}{2}\delta_{x+}^{(\tt l)}\mathcal{I}_{(\tt t)\to (\tt l)}\left(q\mu_{t\cdot}q\right)\nonumber\\
        &- 2\sigma_0^{(\tt l)}\delta_{t\cdot}\zeta + 2\sigma_1^{(\tt l)}\delta_{t\cdot}\delta_{xx}^{(\tt l)}\zeta \label{eqn: fd-z-lossy}
    \end{align}
\end{subequations}
In the context of an implicit scheme (or $\theta$-scheme), we introduce the free parameter $\theta=(1+4/\pi^2)/2\geq1/2$, for enhanced numerical accuracy, as advocated in \cite{verfurth1998posteriori,chaigne1992use}.
$\mu_{x\cdot}$ denotes an operator that takes an average of spatially neighboring values.
Regarding spatial resolution, we determine $h^{(\tt t)}=\sqrt{(\gamma^2k^2+\sqrt{\gamma^4k^4+16\kappa^2k^2})/2}$ for transverse directions and $h^{(\tt l)}=\gamma\alpha k$ for longitudinal directions.

\begin{figure}[t]
    \begin{minipage}[b]{0.5\linewidth}
        \centering
        \centerline{\includegraphics[width=\linewidth]{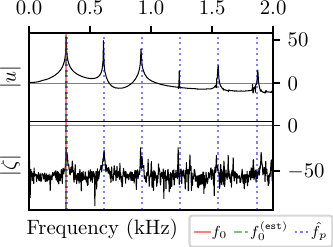}}
        \centerline{(a) $\kappa=6.08$, $c_0=0.0078$}
    \end{minipage}
    \begin{minipage}[b]{0.5\linewidth}
        \centering
        \centerline{\includegraphics[width=\linewidth]{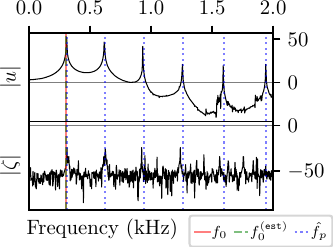}}
        \centerline{(b) $\kappa=9.63$, $c_0=0.0085$}
    \end{minipage}
    \vspace{-6mm}
    \caption{Log-magnitude spectrum of plucked $u$ and $\zeta$ in dB scale.}
    \label{fig: spectrum}
    \vspace{-3mm}
\end{figure}

\subsection{Coupling with excitations}
\label{ssec: coupling with excitations}
\vspace{-1mm}
In this paper, we explore three distinct methods of exciting musical strings: pluck, bow, and hammer.
While implementing a pluck involves setting an initial condition, simulating bow and hammer excitations necessitates introducing additional terms represented as $\Gamma_{\tt B}$ and $\Gamma_{\tt H}$, respectively, to reflect the excitations on $u$.
To model these excitations, we augment eq \eqref{eqn: fd-u-lossy} by $\Gamma_{\tt B}$ and $\Gamma_{\tt H}$ with approximating their forcing factor as part of a lumped mass-spring system.
This section introduces how these are jointly determined along with $u$.

\vspace{-2mm}
\subsubsection{Bow model}\label{sssec: bow}
\vspace{-1mm}
The system of the bow acting on an ideal string is described as:
\begin{subequations}\label{eqn: bow}
    \begin{align}
        -\Gamma_{\tt B} &= J_p(x_{\tt B})F_{\tt B}\varphi(v_{\tt rel}) \label{eqn: bow-gamma}\\
        v_{\tt rel} &= I_p(x_{\tt B})\delta_{t\cdot}u - v_{\tt B} \label{eqn: bow-vrel}
    \end{align}
\end{subequations}
\setlength{\intextsep}{2.5mm}%
\setlength{\columnsep}{8pt}%
\begin{wrapfigure}{r}{0.24\linewidth}
    \vspace{-3mm}
    \centering
    \includegraphics[width=\linewidth]{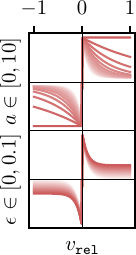}
    \vspace{-7mm}
    \caption{$\varphi$ curves for various $a$, $\epsilon$.}
    \label{fig: friction}
    \vspace{-2.7mm}
\end{wrapfigure}

We have time-varying control parameters: bowing velocity ($v_{\tt B}$), bowing force ($F_{\tt B}$), and bowing location ($x_{\tt B}$).
Additionally, we utilize interpolation and spreading operators ($I_p$ and $J_p$) for position control ($x_{\tt B}$).
Equation \eqref{eqn: bow-vrel} introduces $\delta_{t\cdot}$, implying iterative methods like least squares or the Newton-Raphson method for determining $u(t+1)$.
We employ a hard characteristic curve incorporating sliding friction values, defined as $\varphi(v_{\tt rel}) = \text{sign}(v_{\tt rel})\left(\epsilon+(1-\epsilon)e^{-a|\eta|}\right)$ \cite{bilbao2009numerical}.
While this formula shares some similarities with Woodhouse's characteristic function \cite{woodhouse1992physical}, we introduce randomization to parameters $a$ and $\epsilon$ to diversify bow characteristics.
Figure \ref{fig: friction} illustrates some examples of these characteristic profiles.
Parameter $a$ allows $\varphi$ to deviate from a sign function, while $\epsilon$ serves as an offset to the sliding friction value.

\vspace{-2mm}
\subsubsection{Hammer model}\label{sssec: hammer}
\vspace{-1mm}
Similar to the bow, the system of hammers can be described as: 
\begin{subequations}\label{eqn: hammer}
    \begin{align}
        \Gamma_{\tt H} &= J_p(x_{\tt H})\mathcal{M}F_{\tt H} \label{eqn: hammer-gamma}\\
        F_{\tt H} &= \omega_{\tt H}^{1+\alpha_{\tt H}}\left(\left[u_{\tt H} - I_p(x_{\tt H})u\right]^{+}\right)^{\alpha_{\tt H}-1}\mu_{t\cdot}\left(u_{\tt H} - I_p(x_{\tt H})u\right)\label{eqn: hammer-F}
    \end{align}
\end{subequations}
In the notation, $[\cdot]^+=\max(\cdot, 0)$ represents the ReLU function.
The presence of $\mu_{t\cdot}$ in eq \eqref{eqn: hammer-F} necessitates the utilization of an iterative solver for determining $u_{\tt H}$ and $F_{\tt H}$.
With given initial displacement and velocity, the hammer is approximated by a lumped mass interacting with $u$.
Some might view the implicit scheme presented here as potentially inefficient.
This perception is not entirely unfounded, as it's possible to formulate the schemes introduced in eq \eqref{eqn: bow} and \eqref{eqn: hammer} as explicit schemes, eliminating the need for iterative solvers.
However, it is essential to recognize that this semi-implicit treatment not only enhances scheme stability but also effectively resolves issues related to numerical dispersion and cutoffs \cite{bilbao2009numerical}.

\subsection{Matrix-vector form}\label{ssec: matrix-vector form}
Spatial recursion can be efficiently implemented through matrix-vector operations, offering particular benefits in scenarios amenable to parallelization with GPUs.
To unify transverse and longitudinal equations into a single vector format, we introduce $\mathbf{w}^n=\left[\mathbf{u}^n, \mathbf{z}^n\right]^\top=\left[u_1^n,...,u_{N^{(\tt t)}-1}^n,\zeta_1^n,...,\zeta_{N^{(\tt l)}-1}^n\right]^\top$, where the superscript $n$ signifies the discrete time step.
Subsequently, eq \eqref{eqn: fd-lossy}, combined with \eqref{eqn: bow} and \eqref{eqn: hammer}, is expressed in matrix-vector form as:
\begin{equation}\label{eqn: vector form}
    \mathbf{A}^n\mathbf{w}^{n+1} + \mathbf{B}^n\mathbf{w}^{n} + \mathbf{C}\mathbf{w}^{n-1} + \mathbf{\Gamma}^n = \mathbf{0}
\end{equation}
Here, $\mathbf{\Gamma}^n = \left[\mathbf{\Gamma}^n_{\tt B}+\mathbf{\Gamma}^n_{\tt H},\mathbf{0}\right]^{\top}$ represents the excitation on $\mathbf{u}$.
It is important to note that $\mathbf{\Gamma}^n_{\tt B}$ and/or $\mathbf{\Gamma}^n_{\tt H}$ can be masked to $\mathbf{0}$ depending on the absence of the corresponding excitation.
Matrices $\mathbf{A}^n$, $\mathbf{B}^n$, and $\mathbf{C}^n$ are defined as:
\begin{subequations}\label{eqn: matrix-ABC}
    \begin{align}
        \mathbf{A}^n &= \left[\begin{array}{cc}\mathbf{Q}^{(\tt t)}_+ +\mathbf{V}^{(\tt t)} & \mathbf{K}^{(\tt tl)} \\ \mathbf{K}^{(\tt lt)} & \mathbf{Q}^{(\tt l)}_+ \end{array}\right], \\
        \mathbf{B}^n &= \left[\begin{array}{cc}-2\mathbf{\Theta} - \mathbf{G}^{(\tt t)} + \mathbf{S}^{(\tt t)} & 2\mathbf{K}^{(\tt tl)} \\ \mathbf{0} & -2\mathbf{I}-\alpha^2\mathbf{G}^{(\tt l)} \end{array}\right], \\
        \mathbf{C}^n &= \left[\begin{array}{cc}\mathbf{Q}^{(\tt t)}_-+\mathbf{V}^{(\tt t)} & \mathbf{K}^{(\tt tl)} \\ \mathbf{K}^{(\tt lt)}  & \mathbf{Q}^{(\tt l)}_- \end{array}\right],
    \end{align}
\end{subequations}
where the block matrix components are defined as follows.
\begin{subequations}\label{eqn: matrix-QKV}
    \begin{align}
        \mathbf{Q}^{(\tt t)}_\pm &= \mathbf{\Theta}^{(\tt t)} \pm2\sigma_0k\mathbf{I}^{(\tt t)} \mp 2\sigma_1k\mathbf{D}_{xx}^{(\tt t)} \\
        \mathbf{Q}^{(\tt l)}_\pm &= (1\pm2\sigma_0k)\mathbf{I}^{(\tt l)} \mp 2\sigma_1k\mathbf{D}_{xx}^{(\tt l)} \\
        \mathbf{K}^{(\tt tl)} &= -\phi^2\mathbf{D}_{x+}^{(\tt t)}\mathbf{\Lambda}^n\mathbf{D}_{x-}^{(\tt t)}\mathcal{I}_{(\tt l)\to (\tt t)} \\
        \mathbf{K}^{(\tt lt)} &= -\phi^2\mathbf{D}_{x+}^{(\tt l)}\mathcal{I}_{(\tt l)\to (\tt t)}\mathbf{\Lambda}^n\mathbf{D}_{x-}^{(\tt t)} \\
        \mathbf{V}^{(\tt i)} &= -\phi^2\mathbf{D}_{x+}^{(\tt i)}\left(\mathbf{\Lambda}^n\right)^2\mathbf{D}_{x-}^{(\tt i)} \\
        \mathbf{\Lambda}^n &= \text{diag}(\mathbf{D}_{x-}\left[u_1^n,...,u_{N-1}^n\right]^\top) \\
        \mathbf{G}^{(\tt i)} &= \gamma^2k^2\mathbf{D}_{xx}^{(\tt i)} \\
        \mathbf{S}^{(\tt i)} &= \kappa^2 k^2\mathbf{D}_{xxxx}^{(\tt i)} \\
        \mathbf{\Theta}^{(\tt t)} &= \theta\mathbf{I}^{(\tt t)} + (1-\theta)\mathbf{M}_{x\cdot}^{(\tt t)}        
    \end{align}
\end{subequations}
Note that $\mathbf{I}^{(\tt i)}$, $\mathbf{M}^{(\tt i)}$, and $\mathbf{D}^{(\tt i)}$ represent $(N^{(\tt i)}-1)\times(N^{(\tt i)}-1)$ square matrices corresponding to identity, $\mu$, and $\delta$ operators for $(\tt i)=(\tt t),(\tt l)$.
The parameter $\phi^2$ denotes $\gamma^2k^2(\alpha^2-1)/4$.
For a given set of initial conditions $\mathbf{w}^0$ and $\mathbf{w}^1$, the FDTD simulation recursively computes $\mathbf{w}^{n+1}$ for $n=1,2,\cdots$.
As elucidated in section \ref{ssec: coupling with excitations}, both $\mathbf{\Gamma}^n$ and $\mathbf{w}^{n+1}$ are jointly determined through an iterative method.

\subsection{Implementation details}
We have implemented the simulation described by equation \eqref{eqn: matrix-ABC} using a combination of Python and C++.
Specifically, the Python front-end facilitates easy interaction, while the core FDTD computations are executed in C++.
We've leveraged the PyTorch package to provide flexible hardware acceleration options, abstracting the computing device using the PyTorch C++ extension with ATen.
This abstraction ensures that code initially designed for CPUs can seamlessly run on GPUs, where the C++ operations are automatically dispatched to GPU-optimized implementations.
While further performance enhancements may be attainable through CUDA implementations to fully harness GPU parallelism, their impact may be limited, particularly in cases involving nonlinear recurrent systems like finite difference schemes applied to planar string vibrations.

\begin{figure}
    \centering
    \includegraphics[width=\linewidth]{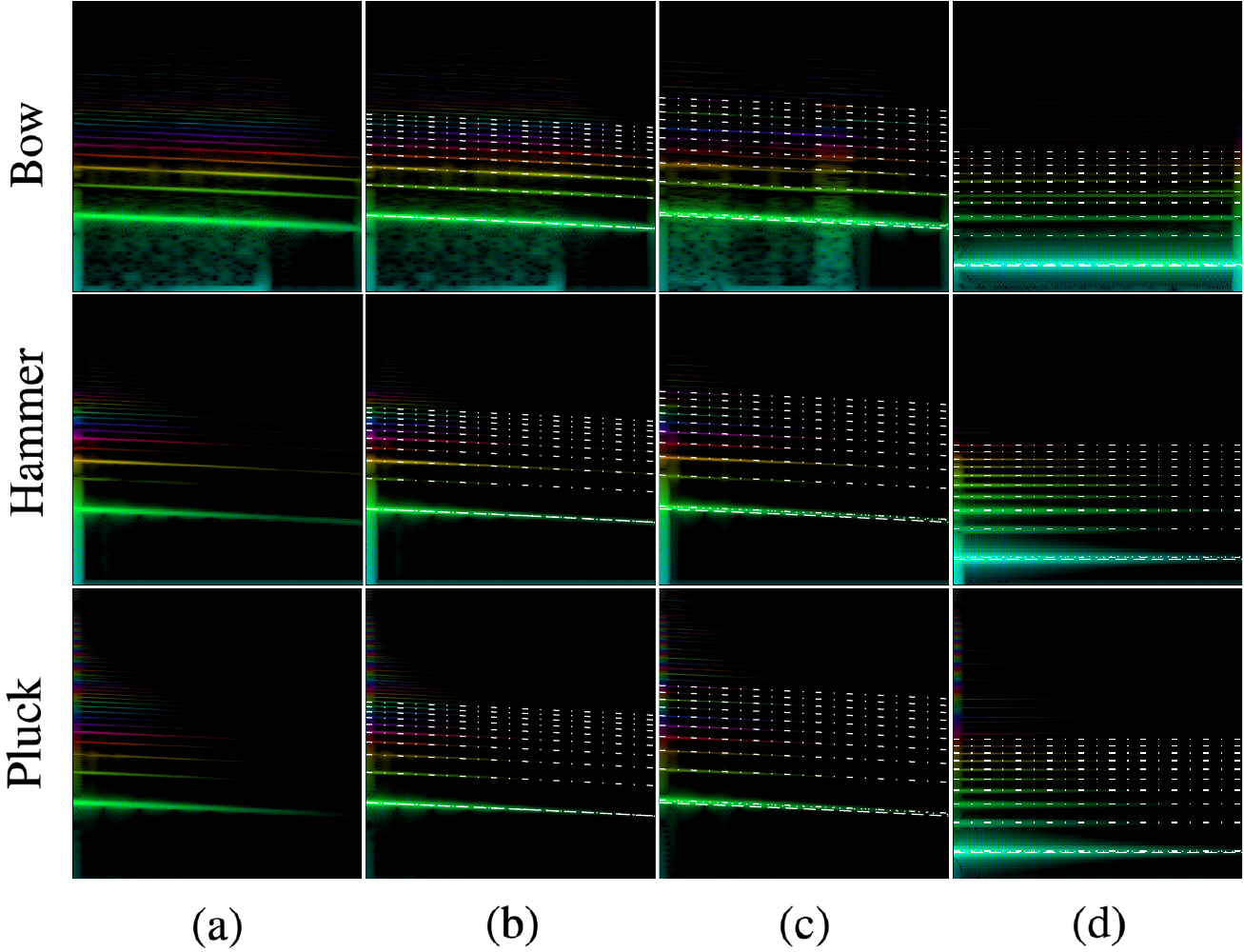}
    \vspace{-6.5mm}
    \caption{Spectrograms of simulated lossy and nonlinear stiff strings. From the 2/3rd point onwards, the bowing force is deliberately adjusted to zero. The overlays of white lines indicate: input $f_0$ (\dashed), estimated $f_0^{(\tt est)}$ (\dotted), and modes $\hat{f_p}$ (\dashdot, $p=0,1,\cdots,9$).}
    \label{fig: rainbowgram}
    \vspace{-4mm}
\end{figure}

\section{Analysis}
\label{sec: analysis}
\vspace{-2
mm}

\subsection{Modal analysis}
\label{ssec: modal analysis}
\setlength{\intextsep}{5pt}%
\setlength{\columnsep}{8pt}%
\begin{wrapfigure}{r}{0.4\linewidth}
    \vspace{-5.5mm}
    \centering
    \includegraphics[width=\linewidth]{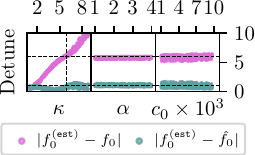}
    \vspace{-6mm}
    \caption{Detune in Hz.}
    \label{fig: kappa-detune}
    \vspace{-1mm}
\end{wrapfigure}

When dealing with a small $\kappa$, encountered in most strings, it is common to express frequencies in terms of the fundamental frequency of ideal lossless strings, $f_0 = \gamma /2$.
This is done using an inharmonicity factor $K=\left(\pi\kappa/\gamma\right)^2$.
We follow the modal analysis established by Fletcher \cite{fletcher1964normal}.
The $p$-th mode for a lossless stiff string with a clamped boundary condition is expected as follows.
\vspace{-1mm}
\begin{equation}\label{eqn: mode}
    \hat{f_p}=f_0(p+1)\left(1 + \frac{2}{\pi}K^{\frac{1}{2}} + \frac{4}{\pi^2}K\right)\left(1+K(p+1)^2\right)^{\frac{1}{2}}
\end{equation}
\vspace{-1mm}
To estimate fundamental frequencies in the synthesized sound, we utilize CREPE \cite{kim2018crepe} and denote the results as $f_0^{(\tt est)}$.

Figure \ref{fig: kappa-detune} shows the scatter plot of the difference in fundamental frequency, or detune, in the simulated pluck sound.
Here, we disregard dissipation by setting $\sigma_0=\sigma_1=0$.
In line with eq \eqref{eqn: mode}, the detuning between $f_0^{(\tt est)}$ and the input $f_0$ increases as $\kappa$ grows.
Furthermore, the simulated $f_0^{(\tt est)}$ exhibits minimal discrepancies (less than 2 Hz) when compared to the first mode $\hat{f_0}$ obtained by eq \eqref{eqn: mode}, even when subjected to changes in stiffness ($\kappa$).
As illustrated in Figure \ref{fig: spectrum} and \ref{fig: rainbowgram}, the simulated harmonic structure also agrees well with the higher modes $p\geq1$.
Nevertheless, setting $\alpha=1$ nullifies the nonlinear coupling term, transforming the system into a linear one.
Yet, Figure \ref{fig: kappa-detune} demonstrates that $f_0^{(\tt est)}$ agrees well with the mode in eq \eqref{eqn: mode} even when varying $\alpha\in[1, 4]$ and $c_0\in[10^{-3}, 10^{-2}]$ (with $\kappa$ fixed at 5.88, as indicated by the dashed line).

\vspace{-3mm}
\subsection{Nonlinearity}
\label{ssec: nonlinearity}
\vspace{-1mm}
Figure \ref{fig: spectrum} shows a log magnitude spectrum of plucked stiff string.
Our chosen parameters for this scenario are: $f_0=300$ Hz, $\alpha=3$, and $x_{\tt P}=0.14$.
Examining the spectra of the displacement variable $u$, we observe prominent harmonics near the modes, accompanied by phantom partials in the high-frequency regions.
Similarly, the spectra of the variable $\zeta$ exhibit peaks, albeit with significantly lower amplitudes, as they are influenced by the transverse displacements.

Figure \ref{fig: rainbowgram} depicts spectrograms of the generated samples with frequency axes displayed in a logarithmic scale.
These spectrograms employ a rainbow colormap, where color indicates instantaneous frequencies and color intensity conveys the log magnitude of the power spectra, as established in \cite{engel2017neural}.
Panel (a) mirrors the spectrograms in (b) but omits the $f_0$ overlays.
The stiffness-to-tension ratio $\alpha$ is consistently set to 1.56 for panels (b-c) and 2.12 for (d).
For (b), we choose a stiffness $\kappa=0.313$, while (c-d) employ $\kappa=9.40$.
As for the increment in $\kappa$, the frequencies in modes are higher in (c) compared to (b), and the spectrogram shows that the simulation results also follow them well.
Although the phantom partials are observable in (c), panel (d) exhibits even stronger intensity.

\begin{figure}
    \vspace{-5mm}
    \centering
    \includegraphics[width=\linewidth]{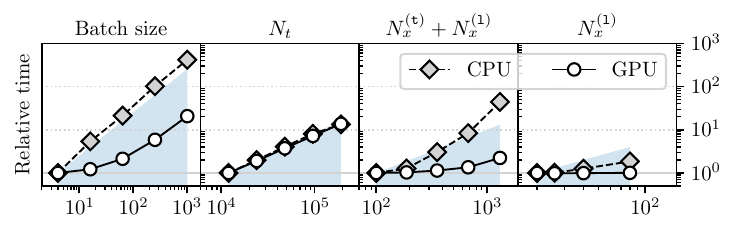}
    \vspace{-8mm}
    \caption{Log-log plot of computational time ratio against grid sizes.}
    \label{fig: time}
    \vspace{-5mm}
\end{figure}

\vspace{-3mm}
\subsection{Computational costs}
\label{ssec: computational costs}
\vspace{-1mm}
Figure \ref{fig: time} illustrates a comparative analysis of computation time increments associated with batch size, and temporal ($N_t$), transverse ($N_x^{(\tt t)}$), and longitudinal ($N_x^{(\tt l)}$) grid size.
The experiment employs either a single GPU (RTX 2080) or a single CPU thread (Xeon E5 v4 @2.1GHz).
We conduct 100 repeated measurements for each configuration.
Shaded regions highlight the sublinear regime, where the time increment rate is less than or equal to the grid increment rate.

Three key observations regarding computational cost are worth noting:
(1) Batching significantly enhances data generation efficiency.
While CPU time scales linearly with batch size, GPU time exhibits sublinear behavior.
In particular, GPUs with a batch size of 1024 operate nearly 50 times faster than those with a batch size of 4.
(2) Recurrence in $N_t$ poses a substantial bottleneck.
Both CPU and GPU times increase proportionally with temporal length, which is unsurprising given that temporal recurrence constitutes the heaviest and irreducible \texttt{\footnotesize for-loop} in the entire process.
The GPU time of simulating 1-sec in $f_s=48$ kHz with batch size 1 takes approximately 17 mins.
(3) Enhancing CPU performance can be achieved by selecting the smallest grid size within the range corresponding to the $f_0$ to be simulated.
While the grid size $N_x$ may determined based on the $f_0$ and $\alpha$ values, it is acceptable to use $N_x$ values larger than the predefined grid size through zero-masking.
Nevertheless, Figure \ref{fig: time} strongly recommends utilizing the minimum $N_x$, particularly when simulating using CPUs.

\vspace{-3mm}
\section{Conclusion}
\label{sec: conclusion}
\vspace{-2mm}
This paper introduces a versatile nonlinear string sound synthesizer capable of simulating planar string motion under various excitations.
Simulator, equipped with the ability to randomize parameters, stands as a valuable resource for the audio signal processing community.
It also offers substantial support for the domain of neural network-based audio synthesis by serving as a high-fidelity dataset constructor.
Its PyTorch-based implementation further enhances its flexibility, making it accessible for both CPU and GPU utilization.

\vspace{-3mm}
\section{Acknowledgement}
\vspace{-2mm}
This work was partly supported by Institute of Information \& Communications Technology Planning \& Evaluation (IITP) grant funded by the Korean Government 2022-0-00641 (50\%) and
National Research Foundation of Korea (NRF) grant funded by the Korea government (MSIT) (No. RS-2023-00219429) (50\%).

\vfill\newpage
\bibliographystyle{IEEEbib}
\bibliography{strings}

\end{document}